\documentclass[epj]{svjour}

\usepackage{graphicx}

\begin{document}

\title{Hypernuclear Physics and Compact Stars}

\author{J\"urgen Schaffner-Bielich}

\institute{Institut f\"ur Theoretische Physik, J. W. Goethe
  Universit\"at, D-60438 Frankfurt am Main, Germany}

\date{Received: date / Revised version: date}

\abstract{Hypernuclear physics plays a decisive role for several features
  of compact star physics. I review the impact of hypernuclear
  potential depths, two-body hyperon-nucleon and hyperon three-body
  forces as well as hyperon-hyperon interactions on the maximum mass,
  the mass-radius relation, and cooling properties of neutron stars.
}

\maketitle


\section{Introduction}


One of the most enigmatic and extreme astrophysical objects are
considered to be neutron stars, which are produced in spectacular core
collapse supernova explosions. These compact, massive objects have
typical radii of about 10 km and masses of $1-2M_\odot$. Matter in the
core of neutron stars experiences extreme densities, several times
normal nuclear matter density, i.e.\ $n\gg n_0 = 3\cdot 10^{14}$
g/cm$^3$. The association of supernova explosions and neutron stars
being its remnant is exemplified with the crab nebula which hosts in its
center a pulsar, a rotating neutron star. 

As of today more than 1600 pulsars are known and recorded in the
publicly available pulsar data base at the Australian National Telescope
Facility (ATNF), see \cite{atnf}. The best determined mass is still the
one of the Hulse-Taylor pulsar with $M=(1.4411\pm0.00035)M_\odot$, the
fastest rotating one is the pulsar PSR J1748-2446ad with 716 revolutions
per second. Recently, indications of even more massive neutron stars in
Pulsar--White Dwarfs Systems have been published \cite{Nice:2005fi}.  In
detail, constraints for the masses of four pulsars with a white dwarf
companion have been measured by timing the pulsar signal.  For the
pulsar J0751 +1807 a mass range of $M=2.1\pm 0.2 M_\odot$
($1\sigma$ standard deviation) and $M=1.6-2.5M_\odot$ ($2\sigma$) has
been inferred.

Constraints on the mass and radius of neutron stars can be derived by
observations in the optical as well as in the x-ray band, a booming
field of exploration since the launch of the x-ray satellites Chandra
and XMM-Newton in 1999. The best studied isolated neutron star is RXJ
1856.35-3754, the closest one known. A two-component blackbody fit to the
combined optical and x-ray spectra results in a low soft temperature, so
as not to be in contradiction with the observed x-ray flux. This low
temperature implies a rather large radius, so that the optical flux
comes out right. A conservative lower limit was given in
\cite{Trumper:2003we} as $R_\infty = 16.5$ km (d/117 pc) which would
allow only for extremely stiff equations of state. 

Another way of probing neutron star matter properties is by cooling
observations of supernova remnants, see e.g.\
\cite{Kaplan:2004tm,Kaplan:2006mb}. The observational limits hints at
fast cooling processes in the interior of neutron stars, i.e.\ direct
URCA reactions. Standard conventional cooling curves are too high, so
that either a large nuclear asymmetry energy or strange exotic particles
are needed to generate efficient and fast cooling!

The basic structure of the low-density region of neutron stars is fairly
well-known. The outer crust consists of a lattice of nuclei with free
electrons and is a few 100 meters thick. The sequence of nuclei is
controlled by their binding energies and follows mainly along the
neutron magic numbers 50 and 82 (for a most recent investigation of the
outer crust see \cite{Ruster:2005fm}). Similar features will be discussed
in the context of hypernuclei below. The inner crust starts at the
neutron drip density at $n=4\cdot 10^{11}-10^{14}$ g/cm$^3$ and consists
of a lattice of nuclei with free neutrons and electrons. The core starts
at the end of the inner crust which occurs around half times normal
nuclear matter density.


\section{Hyperons in Neutron Stars!}


The term neutron star implies that the main component of neutron star
matter are just neutrons. However, this picture changes drastically for
matter at extremely high densities, i.e.\ in the core of neutron stars.
Simple arguments for the presence of other more exotic species besides
nucleons, electrons and muons can be given in terms of a free gas of
hadrons and leptons. Matter in $\beta$-equilibrium but with no
interactions starts to populate $\Sigma^-$ hyperons already at 4$n_0$,
where $n_0$ is the normal nuclear matter density, the lighter $\Lambda$
hyperons appear at $8n_0$ \cite{Ambart60}. Inclusion of nuclear forces
generically reduces these critical densities substantially, so that
hyperons appear already around $2n_0$ (see e.g.\ \cite{Sahakian:1963}
and references therein for the very first investigations of this kind).

That interactions are essential for the description of neutron star
properties is evident from the fact that the corresponding equation of
state of a free gas results in a maximum mass of only $M_{\rm max}
\approx 0.7 M_\odot$ (see e.g.\ \cite{OV39}) which is by more than a
factor two smaller than the presently most precisely known pulsar mass
of $1.44 M_\odot$ for the pulsar PSR 1913+16. Hence, effects from strong
interactions are crucial in describing neutron stars raising the maximum
mass from 0.7 to 2 or more solar masses \cite{Cameron59}.
Note, that this is in contrast to white dwarfs which are basically
stabilised by the Fermi pressure of the free electron gas only.

As hyperons are likely to be present in addition to nucleons, one has to
consider the interactions between all stable baryons. Besides the
nuclear force, there is some knowledge from hypernuclear physics about
the interactions between hyperons and nucleons and scarcely between
hyperons themselves. The $\Lambda N$ interactions is very well studied,
the potential depth of $\Lambda$ hyperons is $U_\Lambda = -30$ MeV at
$n=n_0$ (see e.g.\ \cite{Millener88}), so that bound $\Lambda$
hypernuclear states exists. The situation is different for $\Sigma$
hyperons. The only bound $\Sigma$ hypernucleus known so far,
$^4_\Sigma$He, is bound by isospin forces \cite{Hayano89,Nagae98}. A
detailed scan for $\Sigma$ hypernuclear states turned out to give
negative results \cite{Bart99}. The study of $\Sigma^-$ atoms hints at a
sizable repulsive potential in the nuclear core, i.e. at $n=n_0$. On the
other hand, the $\Xi$ nucleon interactions seems to be attractive,
several $\Xi$ hypernuclear states are reported in the literature
\cite{Dover83}.  More recently, quasi-free production of $\Xi$'s reveal
an attractive potential of $U_\Xi=-18$ MeV \cite{Fukuda98,Khaustov2000}
(with relativistic corrections, see \cite{SG00}). Last but not least,
the hyperon-hyperon (YY) interaction is not really well known, there are just a
few double $\Lambda$ hypernuclear events (for a recent review see
\cite{Gal:2003ze}). The interaction between other pairs of hyperons as
$\Lambda\Xi$ or $\Xi\Xi$ is not known at all experimentally. However,
the hyperon potentials are essential for the determination of the
composition of neutron star matter so basic hypernuclear data can
provide substantial input for the modelling of neutron star matter.

Important for the stability of neutron stars is the short-range
repulsion of the baryon-baryon interaction. Fits with nonrelativistic
potentials to $\Lambda$ hypernuclear data show effects from three-body
interactions for the $\Lambda N$ interaction \cite{Millener88}. The
density dependence of the Schr\"odinger equivalent potential is
compatible with the many-body mean-field potential of relativistic
field-theoretical approaches and demonstrates that the hyperon potential
turns repulsive above $2n_0$ \cite{SMB97}. The absence of these
higher-order terms in density is likely to generate too soft an equation of state,
so that the maximum mass of neutron stars falls below the mass limit of
$1.44M_\odot$. Arguably, this might be the reason that modern Brueckner
calculations of neutron star matter with nucleons and hyperons result in
too low neutron star maximum masses. The hyperon three-body force has
not received too much attention recently, but is known for quite some
time to be repulsive in nature for $\Lambda NN$ \cite{Gal:1967} leading
to the needed additional stability for neutron stars.

\begin{figure}
\vspace*{-0.5cm}
\centering{\includegraphics[width=0.9\columnwidth]{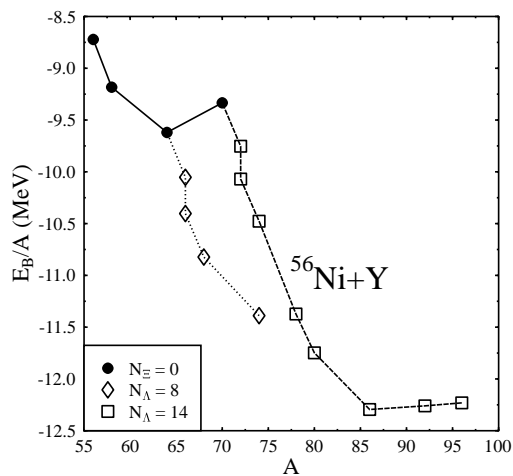}}
\vspace*{-0.5cm}
\caption{The binding energy of strange hadronic matter for a nucleonic
   core of $^{56}$Ni with added $\Lambda$ and $\Xi$ hyperons as a
   function of baryon number $A$ (taken from \cite{Scha93}).}
\label{fig:shm}
\end{figure}

The appearance of hyperons in dense neutron star matter can be also
elucidated by looking at finite systems of nucleons and hyperons, so
called strange hadronic matter \cite{Scha92,Scha93,Scha94,SG00}.  Let us
consider an arbitrary number of nucleons and hyperons forming one big
multi-hypernucleus. The system is stable against strong interactions, if
reactions as $\Lambda + \Lambda \leftrightarrow \Xi + N$ and $\Sigma + N
\rightarrow \Lambda + N$ are Pauli-blocked. The first reaction releases
an energy of $Q\approx 25$ MeV, the second one $Q\approx 80$ MeV which
hints at that $\Sigma$ hyperons can be hardly stabilised in hypernuclear
systems. A similar feature will be present for neutron star matter,
where it is indeed also likely that $\Sigma$ hyperons do not appear
(although the main reason is due to the repulsive potential for $\Sigma$
hyperons). One can construct stable systems of nucleons and hyperons by
adding successively $\Lambda$ hyperons until $\Xi$ hyperons can be
populated as the filled $\Lambda$ hypernuclear levels prevent the strong
reactions by Pauli-blocking. Fig.~\ref{fig:shm} shows the binding energy
of such Pauli-blocked systems for a nucleonic core of $^{56}$Ni versus
the baryon number. When the p-shell of the $\Lambda$ hypernuclear level
is filled up, $\Xi$ hyperons can be added in the s-shell without loosing
stability. On the contrary, the addition of hyperons leads to an overall
increase in the binding energy as the hyperons populate deep lying s--
and p-- states in a separate quantum well.  The nuclear binding energy
with $\Lambda$s and $\Xi$s reaches up to $E/A = -12$ MeV (here a weak
YY interaction is assumed)! In terms of the binding
energy, it is energetically favoured to add hyperons to the system. A
similar effect occurs for dense matter in $\beta$-equilibrium: here
beyond some critical density, the filling of low-lying (with low Fermi
momenta) hyperon states in a newly opened quantum well becomes preferred
compared to adding more nucleons at large Fermi momenta. Hyperons appear
in dense matter when their in-medium energy equals their chemical
potential $\mu(Y) = \omega(Y) = m_Y + U_Y(n)$. Hyperons are then
Pauli-blocked and can not decay as all levels are filled up for its
possible decay products. In the case of neutron star matter, strange
hadronic matter becomes now even stable to weak interactions!

In modern nuclear models, which are fitted to nuclear and hypernuclear
data, hyperons appear in neutron star matter at $n\approx 2n_0$ in
relativistic mean-field (RMF) models \cite{Glen85,Knorren95b,SM96}, in a
nonrelativistic potential model \cite{Balberg97}, in the quark-meson
coupling model \cite{Pal99}, in relativistic Hartree--Fock models
\cite{Huber98}, in Brueckner--Hartree--Fock calculations
\cite{Baldo00,Vidana00}, in chiral effective Lagrangians
\cite{Hanauske00}, in the density-dependent hadron field theory
\cite{Hofmann:2000mc} and in G-matrix calculations
\cite{Nishizaki:2002ih}. It is remarkable that one of the very first
calculations came to a similar conclusion \cite{Sahakian:1963}. Hence,
neutron stars are indeed giant hypernuclei \cite{Glen85}!

\begin{figure}
\centering{\includegraphics[width=0.9\columnwidth]{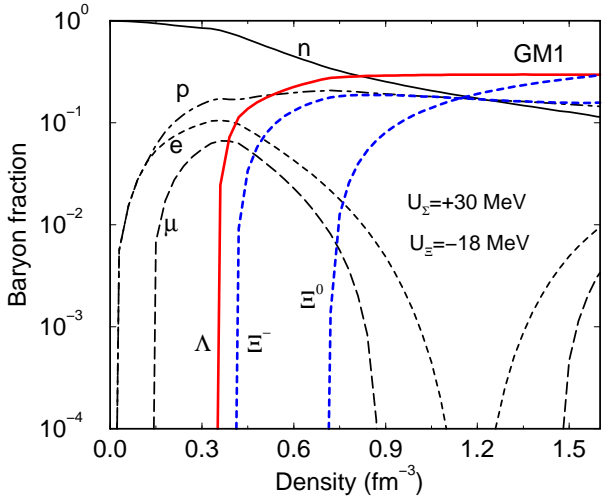}}
\caption{The fraction of baryons and leptons in neutron star matter for
  a RMF calculation using set GM1 with weak hyperon-hyperon interactions
  (see \cite{SM96}).}
\label{fig:comp}
\end{figure}

The composition of neutron star matter depends sensitively on the
assumed hypernuclear potentials. The $\Sigma^-$ hyperon appears in dense
matter usually together with the $\Lambda$ at about $2n_0$, in some
cases even slightly before the $\Lambda$ due to its negative charge, if
an attractive potential of $U_\Sigma=-30$ MeV similar to the $\Lambda$
is chosen. However, for a repulsive potential, the $\Sigma^-$ as well as
the other $\Sigma$ hyperons will not be present in neutron star matter
at all. Fig.~\ref{fig:comp} depicts the fraction of baryons and leptons
as a function of density for a relativistic mean-field calculation using
the parameter set GM1 \cite{GM91} assuming a repulsive $\Sigma$ potential.
The $\Lambda$ is present at $2.3n_0$, the $\Xi^-$ hyperon at $2.7n_0$
(here the model with weak YY interaction is taken from
\cite{SM96}). Besides the $\Xi^0$ emerging at $4.7n_0$ no other hyperon
is present up to $10n_0$, which is well beyond the maximum density
reached for this equation of state. It is clear that hypernuclear data
provides as an essential ingredient the hyperon potential depth which
controls the composition in the core of neutron stars. The baryon and
lepton population is highly sensitive to the in-medium potential of
hyperons which will turn out to be important for the cooling of neutron stars.


\section{Hyperons and cooling of neutron stars}


Moderately aged neutron stars up to 1 million years after their
formation will dominantly cool by volume emission of neutrinos. Cooling
of photons from the surface will take over afterwards. The standard
reaction for cooling is the modified URCA processes $N + p + e^- \to N +
n + \nu_e$ and $N + n \to N + p + e^- + \bar\nu_e$ with a bystander
nucleon to conserve energy and momentum. The modified URCA process is
slow and leaves the neutron star quite warm until the photon cooling
epoch. Much faster reactions are the direct URCA processes as $p + e^-
\to n + \nu_e$ and $n \to p + e^- + \bar\nu_e $. However, this reaction
can only proceed if the Fermi momenta fulfil the condition $p_F^p +
p_F^e \geq p_F^n$. Charge neutrality implies $n_p = n_e$ or $p_F^p =
p_F^e$, so that $2p_F^p = p_F^n$. Hence, the proton fraction has to
exceed $n_p/n \geq 1/9\approx 11\%$ for the direct nucleon URCA process
to start. Relativistic calculations usually reach this value quite
easily \cite{Lattimer:1991ib}. From Fig.~\ref{fig:comp} one can read off
the critical density for the direct nucleon URCA process to be $1.5n_0$.
Nonrelativistic calculations do not get that large proton fraction, as
the asymmetry energy does not have the same strong density dependence as in
relativistic models. In addition, nucleons are pairing strongly, so that
energy is needed to break them up (recent reviews on cooling of neutron
stars can be found in \cite{Yakovlev:2004iq,Sedrakian:2006mq}).

On the other hand, hyperons can help substantially to cool a neutron
star via the hyperon direct URCA processes as $\Lambda \to p + e^- +
\bar\nu_e$ or $\Sigma^- \to \Lambda + e^- + \bar\nu_e$. Remarkably, the
hyperon direct URCA process happen immediately when hyperons are present
and can also occur if there is no direct URCA process for nucleons
allowed \cite{Prakash:1992}! There is no minimum fraction of hyperons needed,
as there is no additional constraint from the charge neutrality
condition as for nucleons (in reality the presence of muons gives a
small critical fraction of a few per mille, see \cite{Prakash:1992}).
Hence, if nucleons are gapped the most important cooling mechanism
involves hyperons. 

For weak YY coupling or interaction strengths, there will be rapid
cooling due to the presence of hyperons mimicking some more exotic agent
as kaon condensation or quark matter in the core. The rapid cooling
process can start basically as soon as hyperons are part of the
composition of neutron star matter, which implies that there is some
critical neutron star mass for fast cooling.  Hyperon cooling is only
suppressed by hyperon pairing gaps which are presumably much smaller
than the ones for nucleons.  Hence, a detailed modelling of the cooling
of neutron stars demands to have a knowledge not only on the
composition, which is fixed by the in-medium potential of hyperons, but
also on the YY interaction strength which determines the hyperon gap
energy. There exist a few studies on hyperon cooling in the literature
(see \cite{Schaab:1998zq,Page:2000wt,Vidana:2004rd,Takatsuka:2005bp} and
references therein). In the first hyperon cooling calculation with
hyperon pairing \cite{Schaab:1998zq}, hyperons are present in the core
for $M \geq 1.35 M_\odot$. The $\Sigma^-$ appears before the $\Lambda$
so that the dominant cooling process involves the reaction $\Sigma^- \to
\Lambda + e^- + \bar\nu_e$. Two-body YY interactions were used as input
to model the hyperon pairing gaps and their emissivities.  It was found
that hyperon gaps improve the thermal history and are more consistent
with x-ray observations of neutron stars.  On the other hand, in a
subsequent study \cite{Page:2000wt} the $\Lambda$ hyperon appeared at a
slightly lower density than the $\Sigma^-$, so that there was a tiny
density range of unpaired $\Lambda$ hyperons present. These unpaired
hyperons resulted in even faster cooling for heavier stars via the
hyperon direct URCA. The conclusion is, that indeed two-body
forces between hyperons and nucleons have an enormous impact on the
cooling history of neutron stars. Hence, hypernuclear physics serves as
a key ingredient not only for the composition of dense neutron star
matter but also for the cooling history of neutron stars.


\section{Hyperons and the maximum mass of neutron stars}


It is known for quite some time, that hyperons have a significant effect
on the global properties of compact stars. As new degree of freedom,
which can populate new Fermi levels, hyperons can lower the overall
Fermi energy and momentum of baryons and leptons. Thereby, the total pressure
of the system for a given energy density is considerably lowered, which
implies that the equation of state is substantially softened.

The first consistent implementation of relativistic hyperon potential
depths in neutron star matter was performed by Glendenning and
Moszkowski \cite{GM91}. Using a relativistic field theoretical approach,
the neutron star with nucleons and leptons only reached a maximum mass
of $M\approx 2.3M_\odot$. A substantial decrease of the maximum mass
occurred once hyperons were taken into account, with parameters fixed by
hypernuclear data. The maximum mass for such ``giant hypernuclei''
turned out to be now around $M\approx 1.7 M_\odot$. Moreover, they
demonstrated that the case of noninteracting hyperons results in a too low
maximum mass, i.e.\ $M<1.4M_\odot$! Clearly, strong (repulsive)
interactions between hyperons have to be implemented for a consistent
description of pulsar masses. 

The issue of the softness of the nuclear equation of state and the
maximum mass of neutron stars has received considerable renewed interest
recently due to the analysis of heavy-ion data. The focus will be here,
in the interest of the present conference, on the analysis of strange
particle production in heavy-ion collisions, in particular the
subthreshold production of kaons measured by the KaoS collaboration
\cite{Forster:2007qk} at GSI, Darmstadt. The analysis of the combined
data with transport models at various collision energies and colliding
systems comes to the conclusion that the nuclear equation of state
should be rather soft at densities around $2-3n_0$
\cite{Hartnack:2005tr}. The important point is that this conclusion is
insensitive to the underlying microphysical input for the transport
simulation, as the kaon-nucleon optical potential, cross sections,
lifetime of resonances etc. The extracted compression modulus turns out
to be around 200 MeV for a simple Skyrme-type parameterisation of the
nuclear equation of state.

However, as outlined above, most recent pulsar data points towards quite
large masses which can be only reconciled with a very stiff hadronic
equation of state. There seems to be an obvious conflict between
heavy-ion data and pulsar observations. The apparent contradiction can
be resolved by noting that transport models use actually the
Schr\"odinger equivalent potential as input not the nuclear equation of
state. Second, the nuclear density ranges probed are different for the
production of kaons and the maximum mass of neutron stars. Typically,
the maximum central density reached in the center of neutron stars
amounts to about say $5-6n_0$, which of course depends on the assumed
hadronic model. These values could be much larger. However, one hardly
finds a calculation in the literature with substantially lower values
for the maximum central densities. As stated above, kaon production in
heavy-ion collisions is sensitive to $2-3n_0$. Therefore, there is a gap
in the nuclear density regions probed. The stiffness of the hadronic
equation of state above $2-3n_0$ controls the value of the maximum mass
achievable for neutron stars. Interestingly, this is the density regime
where hyperons presumably appear and modify the neutron star matter
properties significantly. These lines of arguments have been
cross-checked in a more detailed investigation using Skyrme-type and
relativistic mean-field models which will be reported elsewhere
\cite{Sagert:2007}. The 'soft nuclear equation of state' extracted from
heavy-ion data is indeed compatible with the recent pulsar mass
measurements when only nucleons and leptons are considered as the basic
constituents in neutron star matter.  The inclusion of hyperons,
however, causes an equation of state which turns to be too soft at high
densities with the constraint from heavy-ion data, so that the maximum
mass is lower than the limit from pulsar data!

Again, hyperons play a decisive role in compact star physics. The
feature, that hyperons lower drastically the maximum mass of neutron
stars became even more pronounced with modern many-body approaches to
neutron star matter beyond the mean-field approximation.  In
relativistic Hartree-Fock calculations, the maximum mass of neutron
stars was computed to be $M_{\rm max} = 1.4-1.8 M_\odot$ depending
sensitively on the chosen hyperon coupling strength \cite{Huber:1998hm}.
In Brueckner-Hartree-Fock approaches using Nijmegen soft-core
hyperon-nucleon (YN) potentials maximum masses of $M_{\rm max} = 1.47
M_\odot$ have been derived for the nucleon-nucleon and YN interactions
only and $M_{\rm max} = 1.34 M_\odot$ when including the YY
interactions \cite{Vidana:2000ew}. In the same approach, three-body forces
for nucleons have been included but none for the hyperons so that a
maximum mass of only $M_{\rm max} = 1.26 M_\odot$ was attained
\cite{Baldo:1999rq}. All these mass limits are below the new mass limit of
$1.6 M_\odot$ for the pulsar J0751+1807, the hyperonic equations of
state are just too soft.  Clearly, some additional hyperon physics is
missing.  Presumably, three-body force for hyperons will solve this
problem, as it is repulsive and will raise the maximum mass (some crude
investigations in this directions can be found in \cite{Nishizaki:2002ih}
supporting this statement). Here, input is needed from hypernuclear
physics, not only for the hyperon three-body force but also for the
momentum dependence of the hyperon interactions, as dense matter probes
momenta of the order of several hundred MeVs. Contrary to the widely
used standard mean-field and nonrelativistic approaches, Brueckner-type
approaches adopt momentum-dependent potentials which have to be fixed by
YN scattering and hypernuclear data.

The YY interaction is another important ingredient for the
description of neutron star matter. In fact, it is even possible to
generate a new class of compact stars, hyperon stars, besides ordinary
white dwarfs and neutron stars, by a newly emerging stable solution of
the Tolman-Oppenheimer-Volkoff equation \cite{Scha02}. By increasing the
overall strength of the YY interactions (in particular the
unknown $\Xi\Xi$ interaction which can be probed in heavy-ion collisions
however, see \cite{SMS00}), a first order phase transition appears
from neutron matter to hyperon-rich matter. A mixed phase is present for
a wide range of densities $n_{\rm mix}=(2.5-6.5)n_0$. Interestingly, all
hyperons ($\Lambda$, $\Xi^0$, $\Xi^-$) appear at the start of the mixed
phase, as the bubbles of the new hyperon phase are charged and have a
larger density than the surrounding normal neutron matter (note that for
a Gibbs construction the chemical potentials must be equal in phase
equilibrium, not the densities). The strong first order phase transition
due to hyperons has a strong impact on the mass-radius relation for
compact stars.  A new stable solution in the mass--radius diagram
appears, as the curve reaches a second maximum for the mass for small
radii. Those hyperon stars are generated via attractive YY
interactions (mainly $\Xi\Xi$ interactions) compatible with presently
available hypernuclear data. We note that a weak $\Lambda\Lambda$ does
not rule out a strong $\Xi\Xi$ interaction nor the possible existence of
hyperon stars. The two different solutions behave like neutron star
twins: they have similar maximum masses, $M_{\rm hyp} \sim M_n$, but
different radii $R_{\rm hyp}<R_n$. In addition, selfbound compact stars
for strong YY attraction with $R=7-8$ km are also possible,
but demand that strange hadronic matter is absolutely stable so that
ordinary neutron stars are completely converted to hyperon stars.

Such neutron star twin solutions have been also found for a strong first
order phase transition to quark matter \cite{GK2000,Schertler00,FPS01}.
In fact, any strong first order phase transition can produce a so-called
third family of compact stars. Signals for a such a strong phase
transition can in principle be derived by direct mass and radius
measurements, or by the collapse of a neutron star to the third family
via measurements of gravitational waves, $\gamma$-rays, and neutrinos.

Redshifted spectral lines measured have been claimed to be extracted
from the analysis of x-ray bursts from EXO 0748--676 \cite{Cottam02},
which give a constraint on the mass-radius ratio of the compact star. A
recent analysis of \"Ozel comes to the conclusion that the compact star
mass is $M\geq 2.10\pm 0.28 M_\odot$ with a radius of $R\geq 13.8 \pm
1.8$ km \cite{Ozel:2006km} claiming that 'unconfined quarks do not exist
at the center of neutron stars'! However, this conclusion was put into
perspective in a follow-up reply \cite{Alford:2006vz} which demonstrated
that those limits rule out a soft equations of state, but not quark
stars or hybrid stars. The interactions between quarks can be strongly
repulsive so that the presence of quark matter in the core stabilise the
compact star. On the other hand, the mass limit provides indeed a strong
constraint for hyperons in dense neutron star matter. Hyperons are
likely to appear at moderate densities, which will substantially
decrease the maximum mass. This conclusions is guided by hypernuclear
data and present model calculations. If such massive neutron stars are
confirmed in the future, say with masses above $2M_\odot$, then it seems
that our present understanding of hypernuclear physics of compact stars
will be in conflict with pulsar data!

In passing, I note that strange multiquark states can also exist in
neutron stars, as the H-dibaryon \cite{GS98H} or strange pentaquarks
\cite{Sagert:2006na}. Pentaquarks in neutron star matter will further
reduce the maximum mass, which is being sensitive to the $\Theta^+$
potential. The pentaquark $\Theta^+$ appears around $4n_0$ for a
potential depth of $U(\Theta^+)=-100$ MeV at $n_0$.  For the maximum
mass star the $\Theta^+$ population amounts to 5\% in the core. Present
pulsar mass limits, however, do provide a very weak constraint on
$\Theta^+$ potential (e.g.\ for $M>1.6 M_\odot$, the potential depth
should $U(\Theta^+) > -190$ MeV) which are a much stronger for a
hypothetical negatively charged $\Theta^-$.


\section{Summary}


As outlined above, that hyperons have a substantial impact on
neutron star properties. There is a sizable decrease in the maximum mass
of neutron stars due to the presence of hyperons in the core.  The
$\Lambda$ hyperons appear at $n\approx 2n_0$ in neutron star matter.
The population of $\Sigma$ hyperons hinges crucially on their in-medium
potential. They are likely to be absent for a repulsive potential, but
the negatively charged $\Sigma^-$ could be the first exotic component in
neutron star matter for an attractive potential.  A tiny amount of
hyperons can suffice to cool neutron stars rapidly by the hyperon direct
URCA process, which is controlled by hyperon pairing gaps.  A strongly
attractive YY interaction, between $\Xi$ hyperons, results
in a first order phase transition from neutron-rich to hyperon-rich
matter. This transition allows for a new, stable solution for compact
stars, hyperon stars, with similar masses but smaller radii.  Pulsar
mass measurements can give constraints on multiquark states in dense
matter, e.g.\ for hyponuclei, nuclear systems with a bound
pentaquark state.

It is obvious, that hypernuclear physics provides essential input for
compact star physics. The YN interactions, in particular the
potential depth in bulk nuclear matter, controls the population of
hyperons for massive neutron stars, the first exotic component likely to
appear for supranuclear densities present in the core. The emergence of
hyperons softens the nuclear equation of state and the maximum neutron
star mass possible considerably which depends on the YN
coupling strength and sensitively on the hyperon three-body forces.
Two-body YY interactions regulate the cooling behaviour of
massive neutron stars, as the hyperon direct URCA reaction is suppressed
by hyperon gaps. In addition, hyperons can generate a new class of compact
stars, hyperon stars, for a suitably attractive YY potential. The ongoing
and future experimental hypernuclear programs (see these proceedings) at
DA$\Phi$NE, Jefferson Lab, KEK, J-PARC, MAMI-C, and at GSI, Darmstadt,
in particular the HypHI program and HYPER-GAMMA with PANDA at FAIR, will
provide here the decisive inputs for addressing the global features as
well as the cooling properties of neutron stars.


\bibliographystyle{revtex}
\bibliography{all,literat,hyp2006}

\end{document}